
%
%
\magnification=1200
\hoffset=0.1truein
\voffset=0.1truein
\vsize=23.0truecm
\hsize=16.25truecm
\parskip=0.2truecm
\def\newpage{\vfill\eject}

\def\ep1{ (1 - \epsilon) }
%
%
$\,$
\centerline{\bf A Coupling of Pseudo Nambu Goldstone Bosons}
\centerline {\bf to Other Scalars}
\centerline{\bf and Role in Double Field Inflation}
\vskip 0.2truein
\centerline{\bf Katherine Freese}
\vskip 0.3truein
\centerline{\it $^1$Physics Department, University of Michigan}
\centerline{\it Ann Arbor, MI 48109}
\vskip 0.4truein

\centerline{\it submitted: May 18, 1994}

\bigskip

\medskip
\bigskip
\centerline{\bf Abstract}
\medskip
We find a coupling of Pseudo Nambu Goldstone
bosons (PNGBs) to other (ordinary) scalars, and consider
its importance in various contexts.  Our original
motivation was the model of Double Field inflation [1].
We also briefly consider the role of this coupling
for the case of the QCD axion.


\newpage

The inflationary scenario was proposed in 1981 by Guth [2]
to explain several unresolved problems of the standard
hot Big Bang cosmology, namely the horizon, flatness and
monopole problems.  During the inflationary epoch,
the energy density of the universe is dominated by vacuum
energy, $\rho \sim \rho_{vac}$, and the scale factor
of the universe expands superluminally, $\ddot R > 0$.
In many models the scale factor grows as $R(t) \propto
e^{Ht}$, where the Hubble parameter $H = \dot R/R \sim
(8 \pi G \rho_{vac}/3)^{1/2}$ during inflation.  If the
interval of accelerated expansion satisfies $\Delta t \geq
60 H^{-1}$, a small causally connected region of the universe
grows sufficiently to explain the observed homogeneity
and isotropy of the universe, to dilute any overdensity
of magnetic monopoles, and to flatten the spatial hypersurfaces,
$\Omega \equiv 8 \pi G \rho/3 H^2 \rightarrow 1$.

In the original model of inflation [2], now referred to as
Old Inflation, the universe supercools to a temperature
$T \ll T_c$ during a first-order phase transition
with critical temperature $T_c$.  The nucleation rate
for bubbles of true vacuum must be slow enough that the Universe
remains in the metastable false vacuum long enough for the
required $\sim$ 60 e-foldings of the scale factor.
Unfortunately, the old inflationary scenario has been shown
to fail [3] because the interiors of expanding spherical
bubbles of  true vacuum cannot thermalize properly and produce
a homogeneous radiation dominated universe after the
inflationary epoch.  Subsequently New Inflation [4]
and many other models with rolling fields were proposed,
in which the idea of a strongly first order supercooling
phase transition was abandoned; these new models
were able to successfully reheat back to a radiation
dominated epoch after the inflation.
Extended Inflation[5] and Double Field Inflation[1] were
two different models that were proposed in an attempt
to revive inflation via a first order phase transition.

In the Double Field model, a time-dependent nucleation
rate is obtained via the coupling of two scalar fields.
As shown in [1], the rolling field that catalyzes a time
dependent nucleation rate for the inflaton must have a flat potential
in order to avoid overproducing density fluctuations.
In this paper we investigate using a PNGB for the rolling
field to produce this flat potential (see also [12]).

We have several motivations for considering the Double
Field model with  PNGB field as the rolling field.
This model will produce several features that are different
from what one would obtain from a model
with only one scalar field that is a rolling field
as the inflaton (e.g. a PNGB as inflaton as in Natural
Inflation [6]).
First, the structure produced in such a model
may be different, as there will be bubbles produced
from the first order phase transition as well
as the density fluctuations from the rolling field.
It may be that the spectrum of perturbations from
the rolling field will be different here.  In addition,
at the end of the first order transition, the possibility
exists of producing topological objects such as cosmic strings.
Having inflation take place via a first order phase
transition is simply a different approach from
the purely rolling models.

{\bf Outline:}  In Section I we find an interaction
of a PNGB with other scalars.
We postulate a coupling of the Peccei Quinn
field to another scalar;
at tree level there can be no coupling of the Goldstone
mode (the PNGB) to scalars without violating the underlying symmetry.
However, from the effective action to one loop we find that the
(spatial or temporal) derivative of the goldstone
field couples to the other scalar.
In Section II we investigate the role of this coupling
in Double Field Inflation if the rolling field
is  a PNGB.  In Section III we briefly mention
the role this coupling would have for the case
where the PNGB is a QCD axion.

{\bf  The Term:}
Taking our cue from the axion, we describe a simple
model which has all the ingredients we are interested in.
We consider two scalar fields: a complex field $\Phi_1$
and a real field $\Phi_2$, as well as a fermion $\psi$.
The Lagrangian density is
$$L = |\partial_\mu \Phi_1|^2 + {1 \over 2}(\partial_\mu \Phi_2)^2
- V_{tot}(\Phi_1, \Phi_2) + i \bar \psi \gamma^\mu \partial_\mu \psi
- (h\bar \psi_L \psi_R \Phi_1 + h.c.) \eqno(1)$$
where $\psi_{(R,L)}$ are, respectively, right- and left-
handed projections of the fermion field $\psi_{(R,L)}
= (1 \pm \gamma_5) \psi/2$.  We take the potential
to be
$$V_{tot} = V_1(\Phi_1) + V_2(\Phi_2) + V_I(\Phi_1, \Phi_2) \ .
\eqno(2)$$
The action is invariant
under the global chiral U(1) symmetry:
$$\psi_L \rightarrow e^{i \alpha/2} \psi_L ,
\psi_R \rightarrow e^{-i \alpha/2} \psi_R,
\Phi_1 \rightarrow e^{i \alpha} \Phi_1 \ , \eqno(3)$$
analogous to the Peccei-Quinn symmetry in axion models.
Note that $\Phi_2$ does not transform under the U(1) symmetry.

We take the symmetry to be spontaneously broken at the energy scale $f$,
via a potential
$$V_1(\Phi_1)= \lambda_1\Bigl[\Phi_1^* \Phi_1 - f^2/2\Bigr]^2 \ ,
\eqno(4)$$
where the scalar self-coupling $\lambda_1$ can be of order unity.
Below this scale, the scalar part of the  Lagrangian is then
$$L_{scalar} = |\partial_\mu \Phi_1|^2 + {1 \over 2} (\partial_\mu
\Phi_2)^2 - m_1^2 |\Phi_1|^2 + {1 \over 2} m_2^2 \Phi_2^2
+ V_R(\Phi_1, \Phi_2) \ , \eqno(5)$$
with $- m_1^2 < 0$ since the symmetry is spontaneously broken.
$V_R$ is defined to be all contributions to the potential
other than the mass terms.
We can write
$$\Phi_1 = (\rho + f) e^{i \chi/f} \ . \eqno(6)$$
Below the scale $f$, we can neglect the superheavy radial mode
$\rho$ since it is so massive that it is frozen out,
with mass $m_1 = (2 \lambda)^{1/2} f$.
The remaining light degree of freedom is the angular variable
$\chi$, the Goldstone boson of the spontaneously broken U(1) symmetry
[one can think of this as the angle around the bottom of
the Mexican hat described by Eq. (4)].

Subsequently the symmetry is explicitly broken at at scale
$\Lambda$. For example $U(1)_{PQ}$ is explicitly broken
at the QCD scale $\sim $ 1 GeV, when instantons become
the dominant contribution to the path integral.
Then, for the case of QCD, the Goldstone mode $\chi$ becomes the axion.
In general the potential for the PNGB below
the scale $\Lambda$ becomes
$$V(\chi) = \Lambda^4[1 \pm {\rm{cos}}(N \chi/f)] \ , \eqno(7)$$
where $N$ is the number of minima of the potential.

Because of the $U(1)$ symmetry, the interaction potential
$V_I(\Phi_1, \Phi_2)$ must be independent of $\chi$.
The potential must always be invariant under $\chi \rightarrow
\chi + c$, so that the only coupling we can hope
to get between the Goldstone mode $\chi$ and the field $\Phi_2$
is through derivative couplings involving $\partial_\mu \chi$.
At tree level, the kinetic term in the Lagrangian for the field
$\Phi_1$ is $L_{tree} \sim |\partial_\mu \Phi_1|^2$;
note that this contains a term $\sim (\partial_\mu \chi)^2$.
We will obtain one-loop corrections to this kinetic
term, $L_{1-loop} \sim \bigl[ Z(\Phi_1, \Phi_2) + 1 \bigr]
|\partial_\mu \Phi_1|^2$.
Part of this one-loop correction is the interaction term
we are looking for.

To obtain this we must
calculate the effective $action$ for the field $\phi_1$
as a function of a temporally and or spatially varying
classical goldstone background field.
We follow Ref. [7], who found an effective action
expansion in perturbation theory.
One performs a derivative expansion of the functional determinant
for the gaussian fluctuations around the classical background.
Ref. [7] found the leading order correction to be
$$Z(\Phi_1) = 1 + {1 \over 12 (4 \pi)^2} U^{-1}
\bigl|{\partial U \over \partial \Phi_1}\bigr|^2 \ , \eqno(8)$$
where $U = - m_1^2 + {\partial^2 V_R \over
\partial \Phi_1 \partial \Phi^*_1}$
Again, $V_R$ is defined to be all
contributions to the potential other than the mass terms.
Chan [7] also found higher order terms in the expansion
in
$\partial_\mu \chi \over f^2$; we will assume that this ratio
is small so that we can ignore
higher order derivatives and consider only the leading term above.

For example, we consider
$V_R = {1 \over 4} \lambda_1 |\Phi_1|^4 + {1 \over 24} \tilde \lambda_2
\Phi_2^4
+ \lambda_3 |\Phi_1|^2 \Phi_2^2$.
Then $U = -m_1^2 + \lambda_1 \Phi_1^2 + \lambda_3 \Phi_2^2$
and
$$Z(\Phi_1) = 1 + {1 \over 12 (4 \pi)^2} [-m_1^2 + \lambda_1 |\Phi_1|^2
+ \lambda_3 \Phi_2^2]^{-1} \lambda _1^2 |\Phi_1|^2 \ . \eqno(9)$$
Thus, to one loop, our interaction term is
$$V_I(\chi, \Phi_2) =
{(\partial_\mu \chi)^2 \lambda_1^2 |\Phi_1|^2
\over [-m_1^2 + \lambda_1 |\Phi_1|^2 + \lambda_3 \Phi_2^2]
12 (16 \pi^2)} \ . \eqno(10)$$

{\bf Role of the Coupling in Double Field Inflation:}
In Double Field Inflation we take the field $\Phi_2$ to
be the inflaton, a field that tunnels from false
to true vacuum via nucleation of bubbles at a first
order phase transition. We take the potential 
to be an asymmetric potential with metastable minimum
$\Phi_- $ and absolute minimum $\Phi_+ $
(see Fig. 1).  The energy difference between the vacua
is $\epsilon$.  In the zero-temperature limit,
the nucleation rate $\Gamma_N$ (per unit time per unit volume)
for producing bubbles of true vacuum in the sea of false
vacuum through quantum tunneling has the form [8]
$$\Gamma_N(t) = A e^{-S_E} \ , \eqno(11)$$
where $S_E$ is the Euclidean
action and $A$ is a determinantal factor which is
generally of order $T_C^4$ (where $T_c$ is the energy
scale of the phase transition).

The basic idea of Double Field Inflation is to have
a time dependent nucleation rate of true vacuum bubbles:
initially the rate is virtually zero, so that the
universe remains in the false vacuum for  a long time
and sufficient inflation takes place.  Then, after
this has taken place, the nucleation rate sharply becomes
very large, so that bubbles of true vacuum nucleate
everywhere at once and the phase transition is able
to complete (unlike in Old Inflation).  The universe
then has a chance to thermalize and return to radiation
domination.  In Double Field Inflation, this sudden
change in the nucleation rate is achieved by the
coupling to a second scalar field, which is  a rolling
field.  The purpose of this rolling field is to serve
as a catalyst for the inflaton to go through the
phase transition.  While the rolling field is near
the top of its potential, the nucleation rate is very small;
once the rolling field nears the bottom, the nucleation rate
becomes very large.  Examples were given in [1], hereafter
Paper I. However, density fluctuations are produced by
quantum fluctuations in the rolling field; the amplitude
of fluctuations is too large (in excess of microwave background
measurements) unless the rolling potential is very flat,
namely the ratio of height to width must be less than
$O(10^{-10})$ [9].  For this reason we have investigated
using a PNGB as the rolling field.  A PNGB can naturally
provide the required flat potential.
Unfortunately, as we show below, this requires the other
potential, the potential for the inflaton, to be flat
as well.

Thus, using the notation above, we take $\Phi_2$ to be
the inflaton, a tunneling field, and the PNGB $\chi$ to be the rolling field.
Then the overall Lagrangian is given by eqn (1) where
the interaction term is given by eqn. (10).
For definiteness, we take the potential of the inflaton
field to be
$$V_2(\Phi_2) = {1 \over 8} \lambda_2(\Phi_2^2 - a^2)^2
- {\epsilon \over 2a} (\Phi_2 -a ) \ . \eqno(12)$$
To leading order, the metastable minimum is given by
$\phi_- = - a$ and the absolute minimum by $\phi_+ = + a$.

For the interaction term we use eqn. (10) above.
We take the spatial derivatives to be zero and focus
on the time derivative $\dot \chi$.  We can take
the amplitude of the rolling field to be $|\Phi_1|^2 = f^2$.
Then the term becomes
$$V_I(\chi,\Phi_2) = {\dot \chi^2 \lambda_1^2 f^2
\over 12 (16 \pi^2) [-m_1^2 + \lambda_1 f^2 +
\lambda_3 \Phi_2^2]} \ . \eqno(13)$$
Here we assume that the field $\Phi_2 \leq |\Phi_1| \sim f$
and expand the denominator (later we will see that
sufficient inflation requires $f \sim m_{pl}$ so
that this assumption is not unreasonable).
Then the coupling part of the term is
$$V_I(\chi,\Phi_2) =
{-1 \over 12 (16 \pi^2)} \dot \chi^2 \lambda_1^2 f^2
\lambda_3 \Phi_2^2 {1 \over (-m_1^2 + \lambda_1 f^2)^2} \ . \eqno(14)$$
As the rolling field $\chi$ rolls down its potential,
the time derivative $\dot \chi$ increases, this coupling
term gets more negative, the entropy $S$ decreases,
and the tunneling rate increases.  Thus the tunneling
rate indeed increases with time, as it must.

The equation of motion for the rolling field is
$\ddot \chi + 3 H \dot \chi = - {dV \over d\chi}$;
we will assume the slow roll limit in which the $\ddot \chi$
term is negligible.  Then, using $V(\chi)$ from eqn. (7)
(and assuming the coupling term will not change our result
by many orders of magnitude),  from the equation of motion
we find
$$\dot \chi \sim \Lambda^4 \chi/(3 H f^2) \ . \eqno(15)$$
Here we have made the small angle approximation
sin$(\chi/f) \sim \chi/f$, which is always true
to within an order of magnitude.
We take $H^2 = 8 \pi  M^4/3 m_{pl}^2$ where $M$ is the
energy scale at which inflation takes place and $m_{pl} =
10^{19}$ GeV is the Planck mass.
Thus the term is
$$ V_I(\chi, \Phi_2) = - \eta \chi^2 \Phi_2^2 \eqno(16a)$$ where
$$\eta \simeq 10^{-5}
 \lambda_3 {\Lambda^8 m_{pl}^2 \over M^4 f^6} \ . \eqno(16b)$$

In Paper I we discussed several constraints on the Double Field
model.  We apply these constraints here to the case where
the rolling field is a PNGB with interaction term above,
to find what the factor $\eta$ must be.

First, we want the $\Phi_2$ field to dominate the dynamics
of the Universe and be responsible for the inflationary epoch;
hence we require $V_2(\Phi_2) > V_2(\chi)$, i.e.
$\epsilon \geq \Lambda^4$.

Second, in order for the coupling of the $\chi$ field
to influence the inflaton and bring an end to inflation, we
need  the coupling term to be sufficiently large at the end
of inflation, i.e. $\eta \chi^2 \Phi_2^2 \sim \epsilon$.
Since $\Phi_2 \sim -a$ during inflation, this requirement
becomes $\eta f^2 a^2 \sim \epsilon$.

Third, the slowly rolling field must be able to roll
despite the frictional effect provided by the interaction
term.  We must have $\dot \chi > 0$, i.e.
$\Lambda^4/f^2 - 2 \eta a^2  \geq 0$.

The combination of these three constraints means
that all the terms in the potential must be roughly
comparable,
$$\epsilon \sim \eta f^2 a^2 \sim \Lambda^4 \ . \eqno(17)$$

Bubbles will nucleate at a rate given by Eq. (11).
For the potential of eqn. (13) and in the limit that
the nondegeneracy of the vacua is small (i.e. $\epsilon$ small, thin
wall limit),
the Euclidean action can be obtained analytically [8] and is given
by
$$S_E = {\pi^2 \over 6} {\lambda_2^2 a^{12} \over \epsilon_{eff}^3} \ .
\eqno(18)$$
Here $\epsilon_{eff} = \epsilon + \eta \chi^2 \Phi_2^2 \sim \epsilon$
from above.
To obtain an appropriate nucleation rate, we need $S_E \sim 10$.
Using eqns. (17) and (18) and taking $M \sim \Lambda$,
we find that this requires
$$\lambda_2 \sim 10^{-15} (\Lambda/f)^6 (m_{pl}/f)^6 \ ,
\eqno(19)$$
where we have taken $\lambda_3 = 1$ and $M \sim \Lambda$.
We know [9] that $\Lambda/f \leq 10^{-3}$.
We then have $\lambda_2 \leq 10^{-33} (m_{pl}/f)^6$.
As we have no natural reason to expect otherwise, we would like
to make the coupling $\lambda_2 = 1$ by taking
$f = 3 \times 10^{-6} m_{pl}$ and $\Lambda \sim 3 \times 10^{10}$  GeV.

However, it turns out that scales this low will not allow for
enough inflation in a reasonable fraction of the universe [10].
Here we have the number of e-foldings
$N_T = 3 H^2 \int_{\chi_0}^f{d\chi \over F}$,
where the driving term $F = dV/d\chi = \Lambda^4 {1 \over f}
{\chi \over f} - \eta 2 \chi a^2 \sim \chi \beta \Lambda^4 /f^2$,
where $\beta =O(1)$.
We find $N_T \sim {8 \pi \over \beta}
{f^2 \over m_{pl}^2} {\rm ln}(f/\chi_0)$.
Requiring 60 e-foldings, the value $\lambda_2 =1$ can then
only be obtained if $\chi_0/f = {\rm exp}(-10^{11})$; the field
must start ridiculously near the origin.  The fraction of the
universe which will start out with that value is tiny.

Thus, to obtain a reasonable probability of sufficient
inflation, one must return to the values
$f \sim m_{pl}$ and $\Lambda \sim 10^{-3} m_{pl}$
for the width and height of the rolling PNGB field.
In that case, from eqn. (19) we have the parameter $\lambda_2 \sim
10^{-33}$, a very small number.  Again, one sees the
resurgence of the fine-tuning problem that pervades
many inflationary models.  Of course it may well
be that there will someday be an explanation for this
small number, e.g. it may be related
to the solution to the heirarchy problem.  However,
here we have made it our task to consider a PNGB as
the rolling field in order to try to avoid unexplained
small numbers; yet, a small number surfaced in the potential
for the other scalar field.

We also wish to check the validity of the gradient expansion.
In eqn. (8) we have kept only the first order term and
have neglected terms involving higher derivatives of $\dot \chi$.
This is valid as long as $\dot \chi \ll f^2$.
Using eqn. (15), we see that this roughly requires
$\Lambda^2 {m_{pl} \over f} \ll f^2$.
For $m_{pl} \sim f$ this is definitely satisfied
since $\Lambda /f \leq 10^{-3}$ is required [9].

In short, we have used a PNGB field as the rolling field in the
Double Field Inflation model to naturally explain
the required flatness of the potential for that field.
However, the interaction term that we used in eqn. (16)
leads to a new fine-tuning of the parameters of
the potential for the inflaton field, which is
a tunneling field.   This new small number
arises because the various constraints on
the model require the heights of both potentials
to be comparable and the widths of both potentials
to be comparable.  We suspect that this feature
will be generic to many other attempts to implement
the Double Field model, and that a small parameter
may be difficult to avoid.

{\bf Other Effects:}
The term in eqns. (10) or (16) could serve to couple the QCD axion
to other scalars, and one can check to see if there
are any important effects.  For instance, the other
scalar could be  Higgs or the bosons in non-topological
soliton stars [11]; these scalars could in principle
decay to axions (or to the PNGB in the case of inflation).
However, the decay rates are extremely
suppressed: $V_I \propto {(\partial_\mu \chi)^2 \over m_1^2}
\Phi_2^2 \sim
({m_S \over m_1})^2 \Phi_2^2 \chi^2$,
where $m_S$ is the scalar mass and $m_1$ in the denominator
is the Peccei Quinn scale $\sim 10^{12}$ GeV for the case
of the QCD axion (and $\sim 10^{19}$ GeV for the case of the
inflaton).  Because these couplings are so small we suspect
they are unlikely to be important.

In summary, we found a derivative coupling of a PNGB to other
scalars in eqn. (16).  This coupling could facilitate Double
Field Inflation with a PNGB as the rolling field that
catalyzes a time dependent
nucleation rate for the inflaton.  The potential for the
PNGB is then naturally flat enough to account for
a small amplitude of density fluctuations.
However, we found that the inflaton
potential must also be flat.  We also found that
the same coupling term
in the case of QCD axions is not likely to play a large
role because it is suppressed.

{\bf Acknowledgements:}
I would like to thank Fred Adams, Chris Hill,
Pierre Sikivie, and Rick Watkins for helpful discussions.
I'd especially like to thank Ira Rothstein for
helping me find THE TERM.

\vskip 0.80truein
\centerline{\bf REFERENCES}
\vskip 0.20truein

\item{[1]} F. Adams and K. Freese, {\it Phys. Rev.} D
{\bf 43}, 353 (1991).

\item{[2]} A. H. Guth, {\it Phys. Rev.} D {\bf 23}, 347 (1981).

\item{[3]} A. H. Guth and E. Weinberg, {\it Nucl. Phys.}
{\bf B212}, 321 (1983).

\item{[4]} A. D. Linde, {\it Phys. Lett.} {\bf 108 B}, 389 (1982);
A. Albrecht and P. J. Steinhardt, {\it Phys. Rev. Lett.}
{\bf 48}, 1220 (1982).

\item{[5]} D. La and P. J. Steinhardt, {\it Phys. Rev. Lett.}
{\bf 376}, 62 (1989); {\it Phys. Lett.} {\bf 220 B}, 375 (1989).

\item{[6]} K. Freese, J. A. Frieman, and A.V. Olinto,
{\it Phys. Rev. Lett.} {\bf 65}, 3233 (1990).

\item{[7]} L. Chan, {\it Phys. Rev. Lett.} {\bf 54}, 1222 (1985);
J. Iliopoulos, C. Itzykson, and A. Martin, {\it Rev. Mod. Phys.}
{\bf 47}, 165 (1975).

\item{[8]} S. Coleman, {\it Phys. Rev. D} {\bf 15}, 2929 (1977);
M. B. Voloshin, I. Yu. Kobzarev, and L. B. Okun, {\it Yad. Fiz.}
{\bf 20}, 644 (1975).

\item{[9]}  F. C. Adams, K. Freese, and A. H. Guth,
{\it Phys. Rev.} D {\bf 43}, 965 (1991).

\item{[10]} In [6] we showed that sufficient inflation with only one
scalar field, a PNGB, required $f \sim m_{pl}$.  Here
the other field is responsible for the inflation, yet
if the rolling field is to play an important role in the
nucleation rate, basically it turns out that
the height and width of both
fields must be roughly the same.  In particular,
the number of e-foldings is roughly the same to O(1) as
those calculated for Natural Inflation in [6].

\item{[11]} R Friedberg, TD Lee, and Y Pang, {\it Phys.
Rev. D} {\bf 35}, 3658 (1987);
JA Frieman, GB Gelmini, M Gleiser, and EW Kolb,
{\it Phys. Rev. Lett.} {\bf 60}, 2101 (1988).

\item{[12]} G. Starkman, preprint 1990.

\bye